\documentclass[letterpaper,english,aps, prl, reprint, superscriptaddress]{revtex4-2}

\usepackage[utf8]{inputenc} 
\usepackage[T1]{fontenc}    

\usepackage{amsmath}
\usepackage{amssymb}
\usepackage{bbold}
\usepackage{graphicx}

\makeatletter

\pdfpageheight\paperheight
\pdfpagewidth\paperwidth

\PassOptionsToPackage{position=top}{subfig}
\usepackage{hyperref}
\hypersetup{
	colorlinks = true,
	allcolors = blue
}

\usepackage{times}

\makeatother

\usepackage{babel}

\begin{document}

\title{Entanglement manifestation of knot topology in a non-Hermitian lattice}

\author{Guoying Zhang}
\affiliation{Institute of Theoretical Physics, State Key Laboratory of Quantum Optics Technologies and Devices, Collaborative Innovation Center of Extreme Optics, Shanxi University, Taiyuan 030006, China}

\author{Li Wang}
\email{liwangiphy@sxu.edu.cn}
\affiliation{Institute of Theoretical Physics, State Key Laboratory of Quantum Optics Technologies and Devices, Collaborative Innovation Center of Extreme Optics, Shanxi University, Taiyuan 030006, China}

\author{Shu Chen}
\email{schen@iphy.ac.cn }
\affiliation{Beijing National Laboratory for Condensed Matter Physics, Institute
of Physics, Chinese Academy of Sciences, Beijing 100190, China}
\affiliation{School of Physical Sciences, University of Chinese Academy of Sciences,
Beijing 100049, China }

\date{\today}

\begin{abstract}
Although the homotopy-knot theory has been utilized to implement effective topological classification for non-Hermitian systems,
the physical implications underlying distinct knot topologies remain ambiguous and are 
rarely addressed.
In this work, we propose a one-dimensional non-Hermitian four-band lattice model and 
map out its phase diagram according to the distinct knot structures 
residing in the moment space. The topological phase diagram is ascertained through a spectral winding number.
Furthermore, we derive the exact analytic formula for the phase boundaries that delineate different knot topologies.
To explore the concrete physical implications of distinct knot topologies, we investigate the many-body ground state entanglement entropy for free fermions loaded on such non-Hermitian lattice in real space.
It turns out that different knot topologies imply different magnitudes of entanglement.
Moreover, we show that the central charge $c$ extracted from systematic finite-size scaling of  entanglement entropy provides effective description of the phase diagram of the knot topology.
Finally, we further confirm the phase boundaries for the topological phase transitions alternatively by numerical calculations of the many-body ground state fidelity susceptibility.
Our results showcase the connection between knot topology and entanglement of non-Hermitian systems and may facilitate further exploration of the profound and practical physical implications of knot topology.
\end{abstract}

\maketitle

\textcolor{blue}{\em Introduction.}--
Braiding and knotting, though mundane and ubiquitous in everyday life, are intimately intertwined with sophisticated mathematical theories~\cite{adams2004knotbook}, which have demonstrated inherent relevance and practical applications across many areas of science, from biology~\cite{pnas.1308450110} to physics~\cite{Atiyah1990Geometry,Leach2004,PhysRevLett.111.150404,Pisanty2019,PhysRevLett.122.203201,Witten1989,Lee2020,doi:10.1126/science.aau8740,PhysRevLett.119.147001,PhysRevB.95.094512}.
Mathematically, braids are abstract objects corresponding to interlaced strings in three-dimensional space, and knots can be thought of as closed braids.
Recent theoretical investigations~\cite{PhysRevB.101.205417, prl126.010401, prb103.155129} have revealed that the braid group and knots can more completely and precisely characterize the topology of non-Hermitian periodic systems, in which the braiding operation occurs on complex energy strings.

Non-Hermitian physics~\cite{bender98,RPP70947,NP1411,AP69249,prl121086803,prl121.026808,JPC2035043,PRX8031079,
PRX9041015,PRB99201103,PRL123066404,NP16761,PRL125126402,
PRL124086801,CPB30020506,prl128120401,PRL124056802,PRL125226402,PRL127116801,NC115491,wang2024MEprb,RMP93015005,PRX13021007,flagellateprb} featured complex energies is currently a hot topic that has emerged in recent years, reshaping our understanding of both classical and quantum systems.
It challenges the long-standing paradigm of Hermitian physics, opening doors to a plethora of novel phenomena and concepts.
These include parity-time symmetry~\cite{bender98,RPP70947}, exceptional points (EPs)~\cite{Ozdemir2019,Li18,Parto,PhysRevLett.123.066405}, non-Hermitian skin effect~\cite{prl121086803, Zhang2022}, edge burst~\cite{prl128120401, CPB30020506, PhysRevB.111.174305} and so on.
Correspondingly, topological band theory is extended to non-Hermitian systems~\cite{prl120.146402,RMP93015005}.
Earlier research efforts classified non-Hermitian topology through the perspective of symmetry, which mainly takes root in the gap dichotomy—the distinction between the line gap and the point gap~\cite{PRX8031079,PRX9041015,PRB99235112,PRB100144106}.
Subsequently, homotopy-knot theory~\cite{PhysRevB.101.205417, prl126.010401, prb103.155129} is  proposed for topological classification of non-Hermitian systems primarily based on the observation of braiding and knotting~\cite{PhysRevResearch.4.L022064,PhysRevB.106.L161401,PhysRevLett.130.157201,PhysRevB.110.134317,PhysRevB.106.195425}.
This further strongly fueled the application of braid and knot theory in physics and garnered enthusiastic responses from the experimental community.
Such intriguing topological braiding and knotting of complex energy has been successfully realized and controlled experimentally
via optical fiber resonators~\cite{wangkai2021,PhysRevLett.132.043804}, optomechanical cavities~\cite{Patil2022}, acoustic structures~\cite{PhysRevLett.130.017201,Qiu2023, PhysRevResearch.5.023038,PhysRevLett.132.236401}, nitrogen-vacancy centers~\cite{Yu2022}, ultracold atoms~\cite{PhysRevLett.132.253401}, electric circuits~\cite{PhysRevB.111.075123} and trapped ions~\cite{PhysRevLett.130.163001}.
However, most of the current research efforts are primarily dedicated to exploring the validity of Homotopy-knot theory in achieving a deeper and more comprehensive topological classification of non-Hermitian systems, as well as experimentally demonstrating and verifying of novel braid and knot structures.
The unearthed knot topology in non-Hermitian systems remains an abstract object and the physical implications of knot structures are still rarely addressed and deserve further study~\cite{prl126.010401,PhysRevResearch.4.L022064}, which brings forth the core idea of this work.

In this work, we explore the manifestation of knot topology of non-Hermitian systems via the study of entanglement entropy of free fermions loaded on the lattice. To be
concrete, we construct a one-dimensional fermionic non-Hermitian four-band lattice model.
By tracing the braiding of eigenenergy strings and identifying the resulting knot structure in momentum space, we obtain the knot topological phase diagram of the model.
A variety of braiding behaviors are revealed successively, including unlinks, unknots, Hopf links, and catenanes~\cite{adams2004knotbook}.
Through numerical calculations, we show that the five knot topology phases can be well characterized by a spectral winding number defined on the Bloch Hamiltonian.
Moreover, based on analysis on exceptional points and taking advantage of the analytic eigenenergy formula, we figure out a brief analytical expression for the phase boundaries of the knot topological phase transitions, which is in good agreement with numerical results.
Furthermore,  we try to explore the physical implications of the exhibited knot topology of the model through investigating explicit many-body physical properties of the system in real space.
Utilizing correlation matrix technique~\cite{SciPostPhys.7.5.069, PhysRevResearch.2.033069, Guo_2021}, we investigate the biorthogonal entanglement entropy for free fermions~\cite{Calabrese_2004,Calabrese_2009,SciPostPhys.7.5.069, PhysRevResearch.2.033069, Guo_2021,10.21468/SciPostPhys.11.1.003, Lin2024,cpl_41_12_127302} dwelling on such a four-band lattice at half-filling.
Specific calculation results show that the different knot topologies observed previously imply distinct entanglement properties of the non-Hermitian fermionic many-body system.
As the relevant parameter varies, the entanglement entropy increases or decrease correspondingly.
Further, based on conformal field theory (CFT), we analyze the finite-size scaling behavior of the entanglement entropy, and obtain the central charges $c$ for different knot topological phases through fitting according to the Cardy-Calabrese formula~\cite{Calabrese_2004,Calabrese_2009} for periodic boundary condition (PBC). From the perspective of central charges, the five knot topological phases can be clearly identified, with each knot topological phase possessing a central charge of distinct value.
In addition, we also study the ground state fidelity susceptibilty~\cite{PhysRevLett.96.140604,PhysRevE.74.031123,PhysRevE.76.022101,PhysRevE.76.061108,PhysRevA.77.032111,FidelityGu,PhysRevA.98.052116, Sun2021, PhysRevResearch.3.013015,Tu2023generalpropertiesof} for free fermions in the one-dimensional non-Hermitian lattice.
Clear phase transition boundaries are signaled by divergent behavior of fidelity susceptibilty, which coincide well with the previously obtained analytical result.

\begin{figure}[tp]
\includegraphics[width=7.8cm]{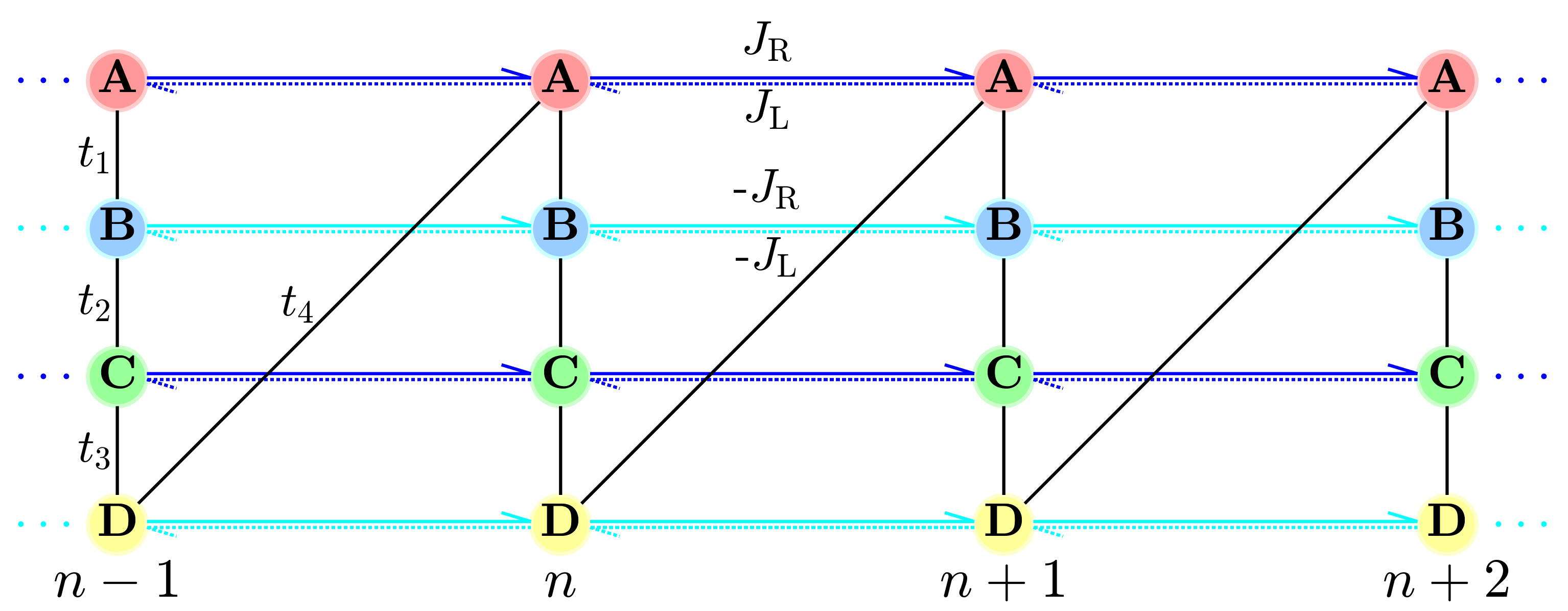}
\caption{ Schematic diagram of the non-Hermitian lattice model with $q=1$. Circles of different colors represent lattice sites of different sublattices, labeled as $A, B, C,$ and $D$. Black solid lines denote normal reciprocal hoppings between different lattice sites, parameterized by $t_1$, $t_2$, $t_3$ and $t_4$,
while blue (cyan) solid and dashed lines with arrows represent non-reciprocal hoppings in two opposite directions, labeled as $J_R$ and $J_L$ ($-J_R$ and $-J_L$),  respectively. $n$ is unit cell index. }
\label{Fig01}
\end{figure}

\begin{figure}[tbp]
\includegraphics[width=8.7cm]{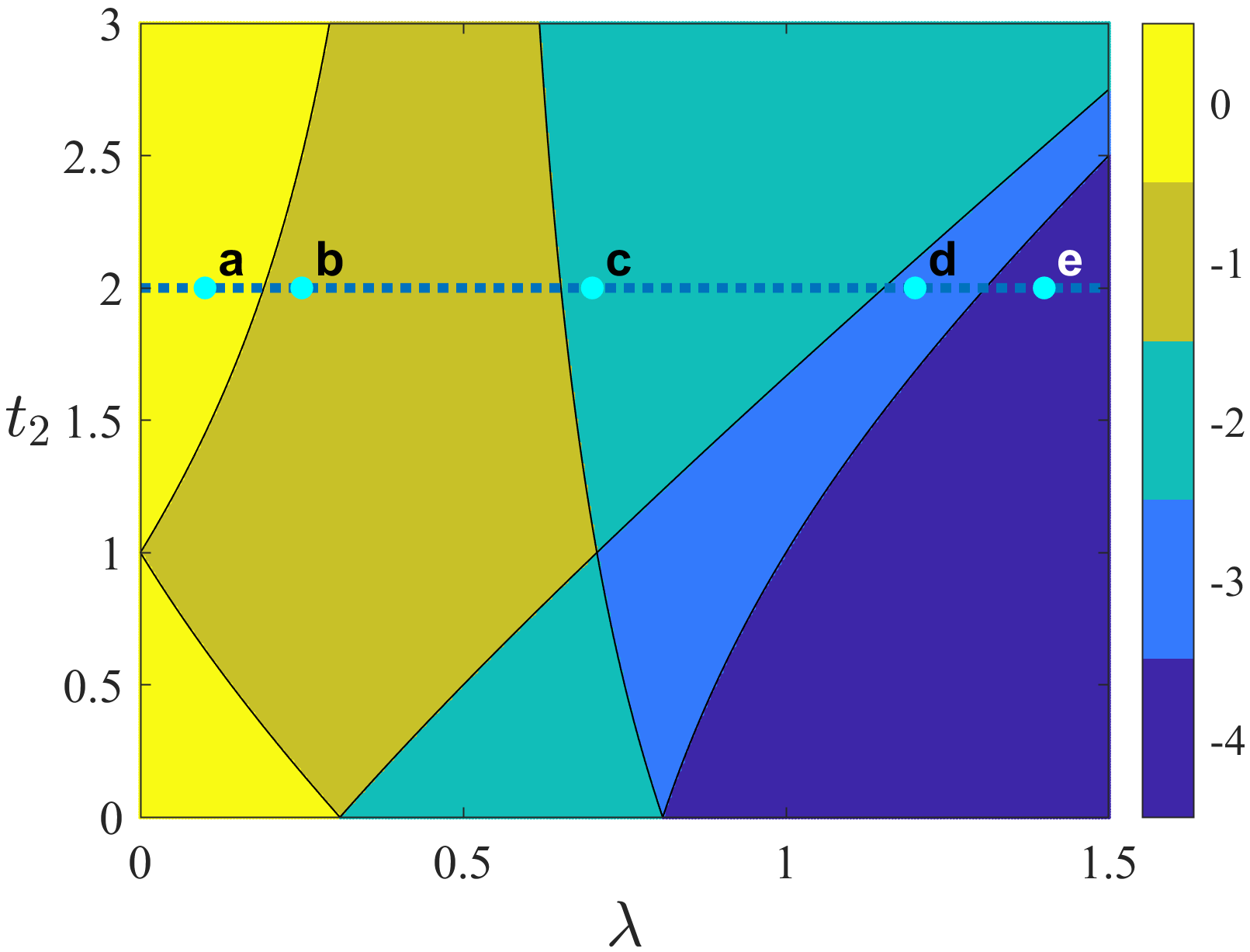}
\caption{
The topological phase diagram of the 1D non-Hermitian system of Eq.~(\ref{Hk}) with $q=1$ in $\lambda\!-\!t_2$ plane.
The phase diagram is partitioned according to the topology of knots formed by the eigenenergy strings of $H(k)$, with each of the five topological phases corresponding to a distinct type of knots.
A representative point is selected for each phase along the line $t_2 = 2$: a ($\lambda = 0.1$), b ($ \lambda = 0.25 $), c ($ \lambda = 0.7 $), d ($ \lambda = 1.2 $), e ($ \lambda = 1.4 $).
The corresponding representative band structures and knots structures of each phase are further shown in Fig.~\ref{knots}.
The colors of different regions in the diagram represent the winding numbers of these regions. The specific correspondence between colors and values can be found in the colorbar. Other parameters are $t_1=t_3=t_4=1, \mu=0.5$.
}\label{Fig02}
\end{figure}

\begin{figure*}[tbp]   
\includegraphics[width=18cm]{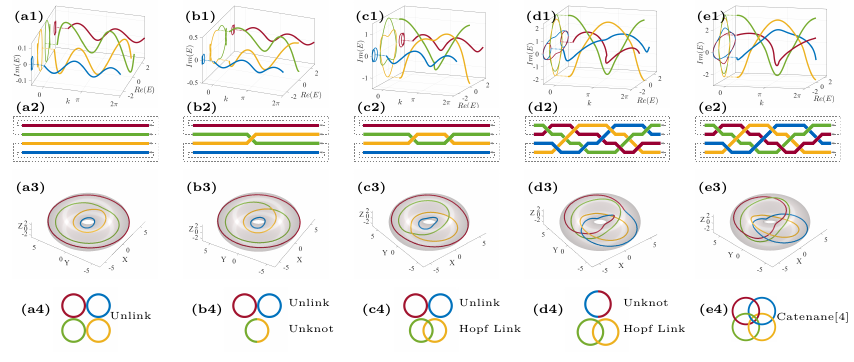}
\caption{\label{knots}
Braiding and knotting of the eigenenergy strings of the model in Eq.~(\ref{Hk}).
Each column corresponds to one of the five marked points (a, b, c, d, e) in Fig.~\ref{Fig02}.
The top row demonstrates typical braid structures formed by eigenenergy strings in the three-dimensional ($\text{Re}(E), \text{Im}(E), k$) space.
The thin solid curves in the $k=-1$ plane are projections of the eigenenergy strings in the complex energy plane.
The second row shows corresponding braid diagrams of the first rows with braid closures by dotted grey lines.
The third row intuitively demonstrates the knots formed by the eigenenergy strings of the one-dimensional non-Hermitian system when the two planes at  $k = 0$ and $k = 2\pi$ are explicitly glued.
The bottom row presents the topological structure of the knots shown above in the third row in a more clear manner.
}\label{Fig03}
\end{figure*}

\textcolor{blue}{\em Model and knot topology}
--We study the knot topologies and their phase transitions in a class of non-Hermitian lattices, as shown in Fig. \ref{Fig01}, which can be modeled by the following tight-binding four-band Hamiltonian,
\begin{align}
H =
& \sum_{n,\alpha} s_{\alpha} \left ( J_L c_{n,\alpha}^{\dagger} c_{n+q,\alpha} + J_R c_{n+q,\alpha}^{\dagger} c_{n,\alpha}  \right) \notag \\
& +\sum_{n} \left ( t_1 c_{n,A}^{\dagger} c_{n,B} + t_2 c_{n,B}^{\dagger} c_{n,C} + t_3 c_{n,C}^{\dagger} c_{n,D} + \right.   \label{Hr}\\
& \left. t_4 c_{n,D}^{\dagger} c_{n+1,A} + h.c \right ) \notag
\end{align}
where $c^{\dagger}_{n,\alpha}$ and $c_{n,\alpha}$ are the creation and annihilation operator of fermions at sublattice site $\alpha$ of unit cell $n$, respectively.
$\alpha$ stands for $A$, $B$, $C$, or $D$.
$s_{\alpha}$ is a sublattice-dependent sign, with $s_A=s_C=1$ and $s_B=s_D=-1$.
$J_R = i(\lambda + \mu)$, and $ J_L = i(\lambda - \mu)$,  describe the hopping amplitudes for right- and left-moving particles within the same sublattice, with $\lambda$ and $\mu$ characterizing non-reciprocal and reciprocal components, respectively.
$q$ is a positive integer that dictates the range of horizontal hopping.
$t_1$, $t_2$, $t_3$, and $t_4$ denote normal reciprocal hoppings between different sublattice sites.

As shown in Fig.~\ref{Fig01}, the system under consideration is imposed with periodic boundary condition (PBC) along the horizontal direction.
Thus, through Fourier transform, one can readily obtain the Hamiltonian of the system in momentum space as follows,
\begin{align}
H(k)=&\frac{t_1 + t_3}{2}I_2 \otimes \sigma_x + \frac{t_1 - t_3}{2}\sigma_z \otimes \sigma_x  \nonumber \\
+& \frac{t_2 + t_4\cos k}{2}\sigma_x \otimes \sigma_x +\frac{t_2 - t_4\cos k}{2}\sigma_y \otimes \sigma_y  \nonumber \\
+&V(k)I_2 \otimes \sigma_z + \frac{t_4 \sin k}{2}(\sigma_y \otimes \sigma_x + \sigma_x \otimes \sigma_y),  \label{Hk}
\end{align}
where
\begin{align}
V(k) = 2\mu \sin(qk) + i 2\lambda \cos(qk),
\end{align}
with $k$  being the wave vector. $I_2$ and $\sigma_{x,y,z}$ are the two-by-two identity matrix and Pauli matrices, respectively.

From the explicit formulation of $H(k)$ given in Eq. (\ref{Hk}), one can see that the non-Hermitian system possesses a generalized particle-hole symmetry (PHS$^{\dagger}$)~\cite{PRX9041015}, represented by $CH(k)^*C=-H(-k)$ with $C=I_2 \otimes \sigma_z$.
Moreover, when $t_1=t_3$ fulfills, the non-Hermitian systems described by Eq.~(\ref{Hk}) respects another symmetry, represented by $T H(k)^{\dagger}T=H(-k)$, where $T=\sigma_x \otimes \sigma_x$.
Furthermore, $H(k)$ with $t_1=t_3$ also has a symmetry described by $\Gamma H(k)^{\dagger} \Gamma^{-1}=-H(k)$ , in which $\Gamma=i(\sigma_x \otimes \sigma_y) K$ with $K$ denoting the complex conjugation operator.
As will be seen later, these symmetries dictate the symmetry of the model's energy spectrum in the complex plane.

The topology of one-dimensional non-Hermitian Hamiltonians with separable energy bands~\cite{prl120.146402} can be completely characterized by the knots (or links) formed by eigenenergy strings~\cite{PhysRevB.101.205417, prl126.010401, prb103.155129} of $H(k)$  as $k$ traverses over a period of $2\pi$.
By numerically solving the Bloch Hamiltonian $H(k)$ in Eq. ~(\ref{Hk}) and tracing the trajectories of the eigenenergies $E_i(k)$ in the ($\text{Re}(E), \text{Im}(E), k$) space, one can obtain the topological phase diagram of the system described by Eq.~(\ref{Hr}) according to knots (or links) classification.
Taking the case with $q=1$ as a concrete example, we demonstrate the topological phase diagram for the non-Hermitian Hamiltonian of Eq.~(\ref{Hk}) in  Fig.~\ref{Fig02} with five topological phases corresponding to distinct knots (links) topologies.
 To avoid verbosity, in what follows,  we will also refer to links loosely as knots.

The specific details of the braiding and knotting of the eigenenergy strings corresponding to each topological phase are displayed in Fig.~\ref{Fig03}.
In  Fig.~\ref{Fig03}(a1-e1), we present the typical structures of braids formed by eigenenergy strings as the wave vector $k$ running from $0$ to $2\pi$.
Projecting all eigenenergy strings onto the  ($\text{Re}(E), \text{Im}(E)$) plane at a fixed $k$ on the left side, one can obtain the complex energy spectrum of the system.
It is evident that under the current parameter settings with $t_1=t_3$, the energy spectrum of the one-dimensional non-Hermitian system is symmetric with respect to both $\text{Re}(E)=0$ and $\text{Im}(E) = 0$, which is a direct manifestation of the symmetry of $H(k)$ that is discussed earlier.
To see more clearly the structures of the braids formed by the eigenenergy strings shown in Fig.~\ref{Fig03}(a1-e1), we explicitly present the corresponding braid diagrams in Fig.~\ref{Fig03}(a2-e2).
Since the two ends of the Brillouin zone (BZ) are equivalent momentum points, one can actually connect them explicitly in a way as depicted by grey dotted line in Fig.~\ref{Fig03}(a2-e2).
Thus, the original ($\text{Re}(E), \text{Im}(E), k$) space turns into a solid torus.
Correspondingly, the eigenenergy strings of the one-dimensional non-Hermitian system running through the solid torus are connected end-to-end, forming knots, which are directly shown in Fig.~\ref{Fig03}(a3-e3).
The topology of the resulting knots can be equivalently and explicitly illustrated via knot diagrams, as presented in Fig.~\ref{Fig03}(a4-e4).
It can be clearly seen that the five topological phases of the one-dimensional non-Hermitian system given by Eq.~(\ref{Hk}) correspond to five distinctly different knot topologies, respectively.

\begin{figure}[tbp]
\includegraphics[width=8.18cm]{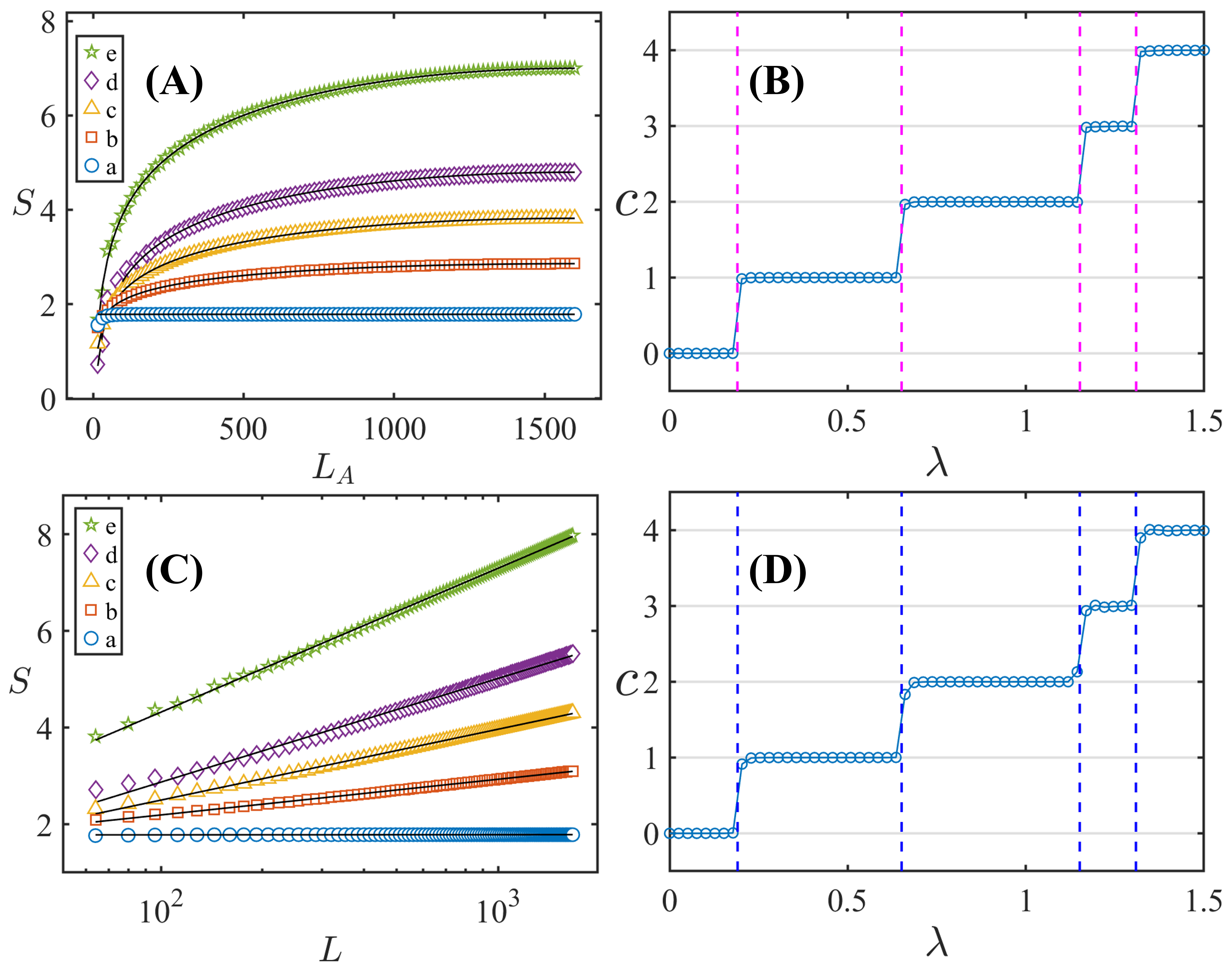}
\caption{
Entanglement characterization of knot topology.
(A) The entanglement entropy $S$ as a function of the subsystem size $L_A$ with the total number of lattice sites as $L=1600$.
Each marker corresponds to an example point selected from different topological phases, as marked in Fig.~\ref{Fig02}.
The black solid line represents the fitting of the entanglement entropy denoted by markers to Eq.~(\ref{SLafit}).
(B) The central charge $c$ extracted from the fitting in (A) as a function of $\lambda$ along the line $t_2=2$ as shown in Fig.~\ref{Fig02}.
(C) The entanglement entropy $S$ as a function of the total number of lattice sites as $L$ with the subsystem size fixed at $L_A=L/2$.
Each marker corresponds to an example point selected from different topological phases, as marked in Fig.~\ref{Fig02}.
The black solid line represents the fitting of the entanglement entropy to Eq.~(\ref{SLfit}).
(D) The central charge $c$ extracted from the fitting in (C) as a function of $\lambda$ along the line $t_2=2$ as shown in Fig.~\ref{Fig02}.
Other parameters are $t_1=t_3=t_4=1, \mu=0.5$.
}\label{Fig04}
\end{figure}

Furthermore, we show that the topology of this one-dimensional non-Hermitian system can be characterized by the winding number $w$, which is defined as~\cite{wangkai2021},
\begin{align}
w =  \oint_{0}^{2\pi} \frac{dk}{2\pi i}\frac{d}{d k} \ln \det\left [ H(k) - \frac{1}{4}Tr[H(k)]\right ] \label{windnum}.
\end{align}
Sweeping across the $\lambda\!\!-\!\!t_2$ parameter space, we calculate the winding number $w$ for each parameter point and marked them with different colors according to their values in the $\lambda\!-\!t_2$  plane.
Evidently,  each topological phase of the one-dimensional non-Hermitian system is characterized by a distinct-valued winding number $w$,  as shown in Fig.~\ref{Fig02}.

On the other hand,  we show that the phase boundaries in the topological phase diagram presented in Fig.~\ref{Fig02} can be given in an analytical manner through meticulous analysis on exceptional points (EPs), as knot topological phase transitions occur via the emergence of EPs~\cite{ prl126.010401}.

Diagonalizing the Hamiltonian $H(k)$  in Eq.~(\ref{Hk}) analytically, one can obtain its four eigenenergies formulated as,
\begin{align}
E(k)=\pm \sqrt{\frac{u}{2}\pm\sqrt{\frac{u^2}{4}-v}+V^2(k)}, \label{Ek}
\end{align}
in which,
\begin{align}
u &= t_1^2 + t_2^2 + t_3^2 + t_4^2, \\
v &= t_1^2t_3^2 + t_2^2t_4^2 - 2t_1t_2t_3t_4 \cos k.
\end{align}

Tedious numerical analysis indicates that the exceptional point only occurs at zero energy.
Thus, taking advantage of the analytical formula Eq. (\ref{Ek}) of the eigenenergies and examining the condition for the occurrence of zero energy degeneracy,
one can accordingly obtain the following analytical formula,
\begin{align}
16\lambda^2 &= 2u \pm \sqrt{4u^2 - 16v}. \label{pb}
\end{align}
When the wave vector $k$ in $v$ explicitly takes values $p\pi/q$ with $p=0,...,q$, Eq.~(\ref{pb}) then yields the phase boundaries that delineate the topological phase diagram.
We have plotted the curves corresponding to Eq.~(\ref{pb}), which are marked with the black solid lines in the Fig.~\ref{Fig02}. It is evident that the curves given by  Eq.~(\ref{pb}) are in good agreement with the boundaries of the topological phase diagram obtained from the previous analysis.

\textcolor{blue}{\em Entanglement manifestation of knot topology in real space}
--Previously, we analyzed the topological phases of this one-dimensional non-Hermitian system primarily in momentum space  based on Eq.~(\ref{Hk}), and
adequately 
characterized them utilizing the abstract concept of braiding and knot topology of the eigenenergy strings.
Now, we investigate the specific physical properties of these topological phases for this one-dimensional non-Hermitian system more intuitively in the real space.
Employing the correlation matrix technique~\cite{SciPostPhys.7.5.069, PhysRevResearch.2.033069, Guo_2021}, we calculated the ground-state bipartite entanglement entropy $S$ of the one-dimensional non-Hermitian lattice given by Eq.~(\ref{Hr}) at half-filling.
To be concrete, the entanglement entropy $S$ of one of the two subsystems (for specificity, labeled as A) can be computed by
\begin{align}
S = -\sum_{m} \left[ \eta_m \log \eta_m + (1 - \eta_m) \log(1 - \eta_m) \right],   \label{SEE}
\end{align}
where $\eta_m$ is the eigenvalue of the correlation matrix $C^A$. The matrix elements $C_{ij}^A$ are defined as~\cite{SciPostPhys.7.5.069, PhysRevResearch.2.033069},
\begin{align}
C_{ij}^{A} = \langle G_{\text{L}} | c_i^{\dagger} c_j | G_{\text{R}} \rangle,    
\end{align}
in which $i$ and $j$ represent lattice sites inside subsystem A.
$ \langle G_{\text{L}} |$ and$ | G_{\text{R}} \rangle$ respectively denote the left and the right fermionic ground state filled by real part of the eigenenergies of the real space Hamiltonian Eq.~(\ref{Hr}).

In Fig.~\ref{Fig04}A, we present the typical entanglement entropies corresponding to each of the five topological phases previously characterized by distinct knot topologies.  $L_A$ denotes the size of the subsystem A.
Representative points taken from different topological phases are denoted by distinct markers.
Different markers correspond to different values of $\lambda$ while $t_2$ is fixed at $2$.
As can be seen from Fig.~\ref{Fig04}A, the entanglement entropies of the five topological phases are distinctly different.
With the increase of $\lambda$, the entanglement entropies increase sequentially in terms of overall profile. 
Furthermore, we find that the scaling behavior of the entanglement entropy $S$ as a function of $L_A$ in the non-interacting non-Hermitian system of Eq.~(\ref{Hr}) still satisfies the Cardy-Calabrese formula~\cite{Calabrese_2004,Calabrese_2009} for PBC as follows,
\begin{align}
S = \frac{c}{3} \log \left[ \frac{L}{\pi} \sin \left( \frac{\pi L_A}{L} \right) \right] + \mathrm{const},  \label{SLafit}
\end{align}
where $L$ denotes the number of total lattice sites and $c$  is the central charge.
Examples of such fits are shown by the solid black lines in  Fig.~\ref{Fig04}A.
The central charges extracted from fittings are demonstrated in Fig.~\ref{Fig04}B.
Interestingly, the five topological phases with different knot topologies exhibit distinct central charges correspondingly.

Moreover, in Fig.~\ref{Fig04}C we further examine the bipartite entanglement entropy $S$ as a function of the total lattice sites $L$.
And to gain an additional perspective, we have adopted a logarithmic scale for the abscissa $L$.
Finite-size scaling of the entanglement entropy shows that the scaling behavior satisfies the following logarithmic function~\cite{Guo_2021},
\begin{align}
S = \frac{c}{3} \log L + \mathrm{const}.   \label{SLfit}
\end{align}
Typical fittings according to the above equation for the five distinct topological phases are demonstrated by the solid black lines in Fig.~\ref{Fig04}C.
Accordingly, the central charges obtained through such fittings are shown in Fig.~\ref{Fig04}D as a function of $\lambda$.
Apparently, they are in good agreement with those presented in Fig.~\ref{Fig04}B.

\begin{figure}[tbp]
\includegraphics[width=8.18cm]{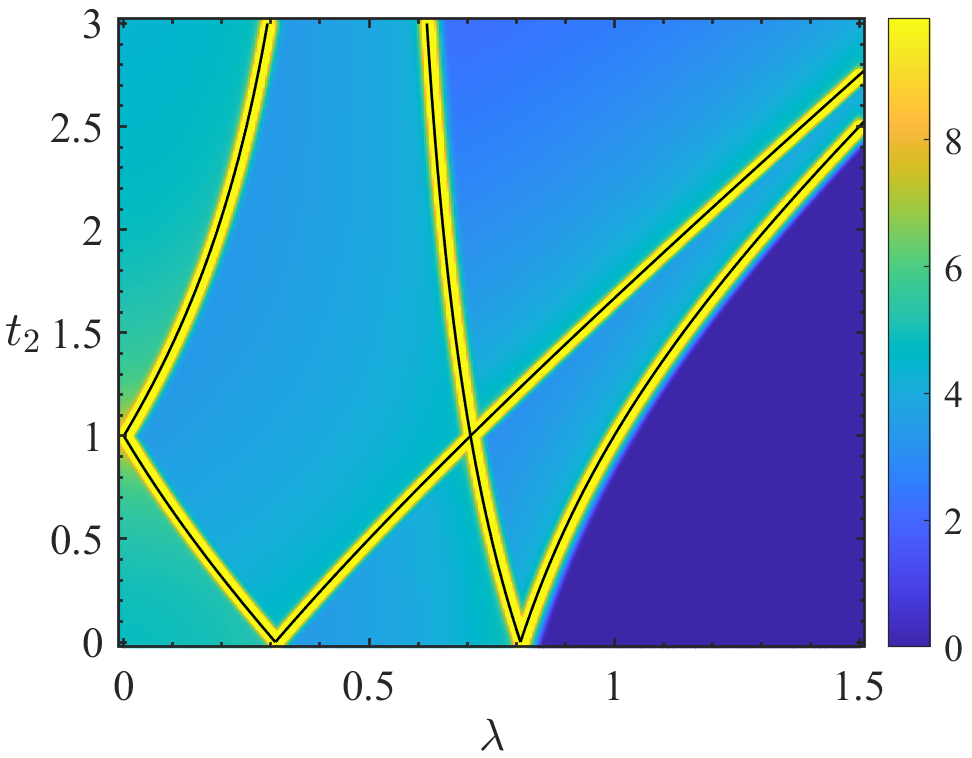}
\caption{
The topological phase boundaries 
identified by the many-body ground state  fidelity susceptibility $\chi_{GS}(\lambda)$.
The color at each parameter point $(\lambda, t_2)$ marks the value of the fidelity susceptibility through function
$\log\left(1 + \vert\chi_{GS}(\lambda)\vert\right)$.
The total number of lattice sites is $L=600$.
The minor perturbation to $\lambda$ is chosen as $\epsilon=0.01$.
To avoid the "smashed out" effect of the maximum value on other sub-maxima and to enable clear visualization of phase boundaries,
the data of $\vert\chi_{GS}(\lambda)\vert$ are clipped at $2\times 10^{4}$.
The black solid lines are plotted according to Eq.~(\ref{pb}).
Other parameters are $t_1=t_3=t_4=1, \mu=0.5$.
}\label{Fig05}
\end{figure}

\textcolor{blue}{\em Topological phase boundaries  confirmation by fidelity susceptibility}
--As is known, fidelity susceptibility~\cite{PhysRevLett.96.140604,PhysRevE.74.031123,PhysRevE.76.022101,PhysRevE.76.061108,PhysRevA.77.032111,FidelityGu,PhysRevA.98.052116, Sun2021, PhysRevResearch.3.013015,Tu2023generalpropertiesof} is a fundamental concept in quantum many-body physics, quantifying the sensitivity of a quantum system to small perturbations in external parameters and serving as a good indicator for identifying quantum phase transitions.
In this section, we apply the many-body ground state fidelity susceptibility~\cite{PhysRevLett.96.140604,PhysRevE.74.031123,PhysRevE.76.022101,PhysRevE.76.061108,PhysRevA.77.032111,FidelityGu,PhysRevA.98.052116, Sun2021, PhysRevResearch.3.013015,Tu2023generalpropertiesof} to identify the topological phase transitions presented in Fig.~\ref{Fig02}.
We adopt the many-body ground state fidelity susceptibility $\chi_{GS}(\lambda)$, which  is defined through the generalized biorthogonal ground state fidelity $F_{GS}(\lambda)$~\cite{PhysRevResearch.3.013015,Tu2023generalpropertiesof} as follows,
\begin{align}
F_{GS}(\lambda) &= \langle G_L(\lambda) | G_R(\lambda + \epsilon) \rangle \langle G_L(\lambda + \epsilon) | G_R(\lambda) \rangle \nonumber \\
&= 1 - \chi_{GS}(\lambda) \epsilon^2 + \mathcal{O}(\epsilon^3),
\end{align}
where $ \langle G_{\text{L}}(\lambda) |$ and $ | G_{\text{R}} (\lambda)\rangle$ respectively represent the left and the right  many-body ground state of the non-Hermitian Hamiltonian in Eq.~(\ref{Hr}) and $\epsilon$ is a sufficiently small number.

Scanning all the parameter points in the $\lambda-t_2$ plane, we numerically calculated the many-body ground state fidelity susceptibility $\chi_{GS}(\lambda)$ of the non-Hermitian fermionic lattice described by Eq.~(\ref{Hr}) at half-filling.
To be concrete, we take the total number of lattice sites as $L=600$, and $\epsilon=0.01$.
As is known, fidelity susceptibility normally will explode around quantum phase transition point, which in turn enables the detection of quantum phase transitions.
In Fig.~\ref{Fig05}, we present the numerical results of the many-body ground state fidelity susceptibility $\chi_{GS}(\lambda)$.
The color of each data point represents the value of $\log\left(1 + \vert\chi_{GS}(\lambda)\vert\right)$.
Apparently, the profile of quantum phase transition boundaries is distinctly manifested in the color pattern labeled according to the values $\chi_{GS}(\lambda)$.
The black solid lines are plotted according to the analytical formula Eq.~(\ref{pb}) for the topological phase transition boundaries.
Obviously, they exhibit good agreement.

\textcolor{blue}{\em Summary.}--
{In this study, we have tried to explore the profound and practical physical implications of intriguing knot topologies through a detailed investigation on a concrete non-Hermitian four-band lattice model.
First,  in momentum space, based on the Bloch Hamiltonian, we analyze the braiding and knotting behavior of its eigenenergy strings  across different parameter regions and demonstrate five distinct knot topological phases. Thus a topological phase diagram is mapped out through knot topology classification.
The five knot topological phases can also be well characterized by a spectral winding number $w$. Furthermore, we manage to obtain the exact analytical formula for the phase boundaries through analysis on the exceptional points of the Bloch Hamiltonian.
Second, in real space, we calculate the many-body ground state entanglement entropy of free fermions for each of the five knot topological phases.
It is shown that different knot topology corresponds to different magnitude of entanglement.
The entanglement entropy can characterize and represent knot topology well.
Specifically, based on conformal field theory and finite-size scaling, we show that different knot topological phases possess distinct central charges $c$, which thus can give out the same topological phase diagram.
Additionally, the phase boundaries delineating distinct knot topologies are further confirmed by numerical calculations on the many-body ground state fidelity susceptibility.
Our findings disclose the correspondence between the abstract mathematical object of  knot topology and the practical physical quantity of entanglement~\cite{Islam2015,Chen2019,Lin2024} for non-Hermitian systems.
This study may further facilitate the exploration of the profound and practical physical implications of the intriguing knot topology.
}

\textcolor{blue}{\em Acknowledgments}--
L.W. is supported by the Fundamental Research Program of Shanxi Province, China (Grant No. 202203021211315), Research Project Supported by Shanxi Scholarship Council of China (Grant No. 2024-004), the National Natural Science Foundation of China (Grant Nos. 11404199, 12147215) and the Fundamental Research Program of Shanxi Province, China (Grant Nos. 1331KSC and 2015021012). S. C. is supported by  by National Key Research and Development Program of China (Grant No. 2023YFA1406704), the NSFC under Grants No. 12174436 and
No. T2121001 and the Strategic Priority Research Program of Chinese Academy of Sciences under Grant No. XDB33000000.

%

\end{document}